\documentstyle[epsf,aasms4]{article}
\begin{document}

\title{Gravitational radiation limit on the spin of young neutron stars}

\lefthead{Andersson et al}
\righthead{The spin of young neutron stars}
\author{Nils Andersson}

\affil{Institut f\"ur Astronomie und Astrophysik, Universit\"at
T\"ubingen, D-72076 T\"ubingen, Germany\\
Department of Mathematics, University of
Southampton, Southampton, UK}

\author{Kostas Kokkotas}

\affil{Department of Physics, Aristotle University of Thessaloniki,
Thessaloniki 54006, Greece}

\author{Bernard F. Schutz}

\affil{Max Planck Institute for Gravitational Physics,
The Albert Einstein Institute,
D-14473 Potsdam, German}

\abstract
A newly discovered instability in rotating neutron stars, driven by 
gravitational radiation reaction acting on the stars' $r$-modes, is shown 
here to set an upper limit on the spin rate of young neutron stars. 
We calculate the timescales for growth of linear perturbations due 
to gravitational radiation reaction, and for dissipation by shear and 
bulk viscosity, working to second order in a slow-rotation expansion 
within a Newtonian polytropic stellar model.  The results are very 
temperature-sensitive: in hot neutron stars ($T>10^9$~K), the lowest-order 
$r$-modes are unstable, while in colder stars they are damped by 
viscosity.  (Similar results have also been obtained independently 
by Lindblom {\em et al} (1998), working however to first order in 
rotation, which omits key terms needed for the viscosity calculation.)  
These calculations have a number of interesting astrophysical implications.
First, the $r$-mode instability will  
spin down a newly born neutron star to a period close to the initial 
period inferred for the Crab pulsar, probably between 10 and 20~ms. 
Second, as an 
initially rapidly rotating star star spins down, an energy equivalent 
to roughly 1\% of a solar mass is radiated as gravitational waves, which makes
the process an interesting source for detectable gravitational waves. 
Third, the $r$-mode instability rules out the scenario whereby millisecond 
pulsars are formed by accretion-induced collapse of
a white dwarf: the new star would be hot enough to spin down to much 
slower rates.  Stars with periods less than perhaps 10~ms must have been 
formed by spin-up through accretion in binary systems, where 
they remain colder than the Eddington temperature of about $10^8$~K.  
More accurate calculations will be required 
to define the limiting spin period more reliably, 
and we discuss the importance of 
the major uncertainties in the stellar models, in 
the initial conditions after 
collapse, and in the physics of cooling, superfluidity and  
the equation of state.  
\endabstract

\section{Introduction}

One of the puzzles of pulsar astrophysics is that neutron stars 
seem to be 
formed with rather small initial spin rates. The best studied 
young pulsar is the Crab (PSR0531+21), whose initial period 
is estimated to have been about 19~ms (Glendenning 1996).  Even 
the recently discovered young 16 ms X-ray pulsar in the supernova 
remnant N157B (Marshall et al 1998) probably had an initial period no 
shorter than 6-9 ms (assuming a braking index typical of young
pulsars). 
This contrasts with the shortest known period 
of a recycled pulsar of 1.6 ms, and with the theoretical lower limit on 
the period of about 0.5 to 2 ms, depending on the equation of state 
(Friedman et al 1989).  If angular momentum is conserved in 
the collapse event that forms the neutron star it would not 
be surprising to find young pulsars spinning at or close to the 
theoretical maximum. 

In this paper we offer an explanation for 
the observed slow initial rotation rates by showing that, if a 
new neutron star has a period shorter than some critical value 
$P_f$, then within a year its period will increase to $P_f$.  During 
this spindown period it will lose angular momentum to 
gravitational radiation, through the growth of a radiation-driven 
nonaxisymmetric instability associated with the $r$-modes.  
The final rotation period $P_f$ 
depends on details of cooling rates, viscosity, and 
perhaps on initial conditions following collapse.
For the simple models of neutron 
stars that we compute in this paper, this period is about 20 ms, 
consistent with the estimated initial spin of the Crab pulsar, 
given the various uncertainties in our calculation (see below).
These uncertainties are, in fact, probably sufficient to 
include the presumed 
initial period of the recently discovered 
pulsar in N157B.  We will discuss 
this further later in the paper. The instability  
operates only in hot neutron stars ($T>10^9$~K); neutron 
stars that are spun up in accreting systems 
do not reach this temperature, and we do not expect the 
instability to put a strong constraint on their speeds. 

An alternative explanation for the slow initial rotation of pulsars
was recently suggested by
Spruit and Phinney (1998).  They argue that pulsars 
are likely to form slowly
rotating because magnetic locking between the core and the envelope of
the progenitor star prevents the core from spinning rapidly.
If this is the case, it has implications for the evolution of 
the angular momentum of the degenerate cores of pre-supernova giants,
and it also reduces expectations that gravitational collapse may 
be followed by a strong burst of gravitational radiation.
However, even if the mechanism suggested by Spruit and Phinney
is operating and explains the slow rotation of most pulsars, 
we believe that  the 16~ms pulsar in N157B is clear
evidence that there is a class of pulsars that form
rapidly spinning (with periods shorter than say 10-20 ms). For
these pulsars the r-mode instability should be relevant, and it is the rotation
rate of such young short-period pulsars that we discuss below.

Besides explaining the slow rotation of young pulsars, the 
$r$-mode instability has another important consequence.  Millisecond 
pulsars (with periods shorter than $P_f$) can only reach their 
rotation speeds by being spun up again after they cool down.  The 
alternative model, that these stars are formed by accretion-induced 
collapse of a white dwarf, is inconsistent with this instability: 
the collapse would form 
a star hot enough to spin down within a year.  So the $r$-mode 
instability gives further weight in support of the 
standard picture that millisecond pulsars are a distinct 
class of recycled pulsars that acquired their angular momentum 
through accretion in binary systems.

The $r$-mode instability is a member of the class of
gravitational-radiation-driven instabilities called CFS instabilities 
(Thorne 1987), after Chandrasekhar (1970), who discovered them, 
and Friedman \& Schutz (1978), who showed that they are generic in 
rotating neutron stars.  The criterion for and mechanism of this 
instability are simple: if the pattern speed of the mode is 
forward-going as seen from far away, but backward-going with respect 
to the rotation of the star, then when the mode radiates away 
angular momentum the star can find a rotation state of lower 
angular momentum {\em and} lower energy.  This allows the mode 
amplitude to grow.  For a mode with angular 
pattern speed $\sigma$ in an ideal inviscid uniformly 
rotating star of angular 
velocity $\Omega$, this means there will be an instability if 
\begin{equation}\label{eqn:criterion}
0<\sigma<\Omega.
\end{equation}
The mechanism is much the same as that 
which drives the Kelvin-Helmholz instability, raising waves 
in the sea (Schutz 1983). Viscosity, however, acts on the mode 
in a different way and, if large enough, will damp out the instability.

Many families of oscillation modes of neutron stars contain members 
that, for sufficiently high stellar rotation rate, satisfy the 
instability criterion (\ref{eqn:criterion}).   Until 
recently it was believed that viscosity in neutron stars stabilized all 
of them except in very limited regimes (Lindblom 1995).  
But the picture changed when Andersson (1998) 
pointed out that a hitherto unstudied family of modes, the 
$r$-modes, which are generalizations of the Rossby waves familiar 
to oceanographers, should also exhibit the instability.  He 
noticed  that,  uniquely among the families of neutron star normal 
modes, the $r$-modes satisfy the instability criterion 
(\ref{eqn:criterion}) for any stellar rotation rate, no matter how 
small. He attempted a preliminary estimate of the growth rate 
of these modes, which showed that it might be interesting and was 
worth further investigation.  Friedman and Morsink 
(1998) then showed that the instability is generic: every $r$-mode 
is in principle unstable in every rotating star, in the absence of 
viscosity. 

To show that the instability is interesting for neutron star evolution, 
however, it is necessary both to calculate accurately the growth rate of 
the mode due to gravitational radiation reaction and to estimate 
the damping caused by viscosity.  In this paper, and in 
independent work by Lindblom et al (1998), it is shown that 
the the $r$-mode instability is surprisingly effective.  In a 
rapidly rotating newly formed star, the mode grows 
rapidly and remains unstable during the time it takes the 
star to cool to about $10^9$~K, about 1 year using conventional 
cooling rates.  During this time
the gravitational radiation emitted by the mode will 
remove most of the star's initial angular momentum, until it 
spins down to roughly the period $P_f$, where viscosity 
dominates gravitational radiation for any temperature.

The $r$-modes are unusual in several respects.  They are primarily 
perturbations of the velocity field in the star, with little disturbance 
in the star's density.  In a nonrotating star, they are neutral
rotational 
motions.  In a rotating star, Coriolis effects provide a weak 
restoring force that gives them genuine dynamics.  The gravitational 
radiation that $r$-modes emit comes primarily from their time-dependent 
{\em mass currents}.  This is the gravitational analogue of 
magnetic multipole radiation.  We believe this instability is 
unique among expected astrophysical sources of gravitational radiation  
in radiating primarily by gravitomagnetic effects, the same 
sector of relativistic gravitation theory that is responsible 
for the Lense-Thirring effect near rotating bodies and the 
Penrose process near spinning black holes (see, for example, Misner et
al 
1973). The present paper and Lindblom et al (1998) give the first 
calculations of the strength of gravitomagnetic radiation from stellar 
perturbations.  

Like magnetic multipole radiation, gravitomagnetic radiation of 
any multipole is $(v/c)^2$ weaker than radiation from the corresponding 
mass (electric) multipole of the same amplitude, where $v$ is the 
typical velocity associated with the perturbation.  Despite this, the 
quadrupole ($l=2$) $r$-mode is more strongly unstable to gravitational 
radiation than any other mode so far examined in neutron stars, such as 
the density-dominated $f$-mode (Lindblom 1995).  The reason 
is that the $r$-modes are unstable in {\em any} inviscid star, while the 
quadrupole $f$-mode is never more than marginally unstable in the most 
rapidly rotating stars, and
the higher-order $f$-modes do not radiate 
any more strongly than the $r$-mode.

Our initial calculation of the strength and the importance of the 
$r$-mode instability necessarily involves approximations.  
The neutron stars are treated as Newtonian polytropes, the 
modes are treated in a slow-rotation approximation, damping 
effects due to viscosity are assumed uniform throughout the 
star, and the stars are assumed to be uniformly rotating.  
An important difference between our calculations and 
the independent ones of Lindblom et al (1998), which were completed 
somewhat earlier, is that we carry the calculation to a sufficiently 
high order in the star's angular velocity $\Omega$ to fully 
calculate the small density perturbation associated with the mass 
motions of the $r$-modes.  This is important for estimating correctly 
the effects of bulk viscosity, which dominates the evolution of the 
spin of the star at high temperatures.  Reassuringly, despite this and 
other smaller differences, the conclusions of the two studies are 
broadly the same.  It seems unlikely to us that when more accurate 
calculations are performed the results will be markedly different 
qualitatively, but there could be quantitative differences in 
the limiting period and in details of the spindown behavior.  

The evolution of the spin of the neutron star and the precise value 
of the limiting period $P_f$ depend also on the physics of neutron-star 
viscosity, cooling, and stratification.  We have adopted the best 
available models of cooling and viscosity (see the references in the 
body of this paper), but there are 
important uncertainties, particularly concerning the onset of 
superfluidity.  We have ignored stratification (differentiation 
of the crust from the interior, for example, or differential rotation 
of the interior).  These uncertainties can also have a significant 
effect on the value of $P_f$.  We hope that the fact that 
$P_f$ is in principle an observable, and that gravitational radiation 
from the spindown may eventually also be directly observed, will 
encourage further work on the critical physics issues here.

In a separate paper (Owen et al 1998), the implications 
of this instability for gravitational radiation are examined.  That 
paper contains a first attempt at developing a time-dependent 
model of the evolution of the instability, including the nonlinear 
phase where it is radiating away most of the angular momentum 
of the star in a steady way.  The conclusions are that (i) 
the radiation emitted by a young neutron star in the Virgo Cluster 
can be directly detected by the advanced versions of the LIGO 
(Abramovici et al 1992) and VIRGO (Bradaschia et al 1990) 
gravitational wave detectors now under construction, and (ii) 
the radiation emitted by all the neutron stars created 
since the beginning of star formation should form a random 
background of gravitational radiation that may be detectable 
by the same instruments.  Such detections could come as early as 
2005.  We will then have direct means of investigating the physics 
that affects the evolution of this instability in neutron stars.

The plan of this paper is as follows.  In Section 2 we describe the 
perturbation formalism and the nature of the $r$-modes, and we
calculate the effects of gravitational radiation, bulk viscosity, 
and shear viscosity on them.  The fundamental limiting period 
$P_f$ is derived in this section.  Section 3 draws our main 
astrophysical conclusions and discusses the chief physics 
uncertainties that affect the computation of the 
limiting period $P_f$. We examine various possible 
spindown scenarios, and identify the key uncertainties that 
can be addressed in future studies.


\section{Estimated timescales}

To investigate the astrophysical relevance of the {\em r}-mode instability
 we have calculated the mode-frequencies and
eigenfunctions from a Newtonian slow-rotation 
expansion accurate to order $\Omega^2$. Our mode-calculation essentially 
repeats
that of Saio (1982) and in order to keep the presentation compact we 
use his notation here. All original equations can be found in his paper.
For the modes that are strongest affected by the instability
(those corresponding to $l=m$) the calculation leads to a displacement vector
\begin{equation}
{\bar \xi}/a =  \left(S_{l+1l} Y_{l+1l},
K_{ll} \sin^{-1}\theta \partial_{\phi} Y_{ll},
          -K_{ll}\partial_\theta Y_{ll} \right) e^{i\omega t} \ ,
\label{rdisp}\end{equation}
where $Y_{lm}$ are the standard spherical harmonics, and
$K_{ll}$  and  $S_{l+1l}$ are eigenfunctions  of order 1 and $\Omega^2$,
respectively.
The variable $a$ is
related to the radial
distance $r$ through $r = a(1+\epsilon)$, where $\epsilon\sim\Omega^2$ is 
the rotationally induced deformation of the star.
We incorporate
these deformations by using a polytropic model created according to 
the procedure of Chandrasekhar and Lebovitz (1962). 
Saio's equations also determine an eigenfunction $\zeta_{l+1l}$ which is
related to the Eulerian perturbation of the pressure through
$\delta p = \rho g a \zeta_{l+1l}Y_{l+1l}$ where   $g=M(a)/a^2$ is the
local acceleration of gravity and
$M(a)$ is the mass inside radius $a$.
  
In our calculation of the r-mode eigenfunctions we make use of 
the Cowling approximation, i.e. we neglect the perturbations
of the gravitational potential. This is not at all necessary, but the
resultant equations are simpler, and it has been shown that
 the error associated with this 
approximation is very small for the {\em r}-modes (Saio 1982; Provost,
Berthomieu
and Rocca 1981). This is natural since the density perturbation
associated with each {\em r}-mode is small.

The character of the {\em r}-modes depends crucially on the stratification
in the star. In consequence, calculations for convective (with
Schwarzschild discriminant $A>0$) and radiative (with $A<0$) lead
to rather different spectra (Provost et al 1981; Saio 1982).
In general, the {\em r}-modes are degenerate at frequency
\begin{equation}
\omega = -m \Omega { (l-1)(l+2) \over l(l+1) }\ ,
\label{rfreq} \end{equation}
measured in the inertial frame, to leading order
(Papaloizou and Pringle 1978).
At higher orders in $\Omega$ the degeneracy of the
{\em r}-modes is then broken and modes whose
eigenfunctions behave differently on the radial coordinate $r$ 
(eg. have a different number of nodes) become distinct (Saio 1982).
Thus, an acceptable overall solution to Saio's equations can be found only
for a discrete set of eigenvalues that correspond to small (order $\Omega^3$)
corrections to (\ref{rfreq}). The barotropic perturbations ($A=0$)
that we consider in the present paper are somewhat different.
First of all, we
find no {\em r}-mode solutions for $l\neq m$. This is in accord with the
result of Provost et al (1981). For $l=m$ we find a
single mode-solution to Saio's equation.
The particulars of this solution are easily understood.
In the case of barotropic perturbations Saio's equations simplify
considerably and can be written as
\begin{equation}
a {{d S_{l+1l}}\over {da}} =
  -\left( l+4-{M(a) \over a c^2}\right) S_{l+1l}
  - ( h+ {M(a) \over a c^2})\zeta_{l+1l} \ ,
\label{bspert}\end{equation}
where $c$ is the sound speed, and the variable $h$ depends on the 
rotationally induced deformation of the star (see Saio (1982) for
the definition of $h$).
We also have
\begin{equation}
a{{d\zeta_{l+1l}}\over {da}} =
     \left(l+2- {{4\pi\rho a^3}\over {M(a)}}\right)\zeta_{l+1l} \ .
\label{bzpert}\end{equation}
A mode-solution to these equations is regular at the centre
of the star and satifies the condition that the Lagrangian variation
in the pressure vanishes at the surface. Given such a solution
the toroidal component of the displacement vector follows from the
algebraic relation
\begin{equation}
\zeta_{l+1l} = - {2 i \over \sqrt{2l+3}} {l \over {l+1}} 
(\omega+l\Omega) \Omega \left( {M\over R^3}\right)^{1/2} 
 {a^3 \over M(a)} K_{ll} \ .
\label{kdef}\end{equation}
The barotropic case is peculiar because the equation for
$\zeta_{l+1l}$ is completely decoupled. Furthermore, from
(\ref{bzpert}) and (\ref{kdef}) one can easily deduce that 
the equation allows a solution $K_{ll}=a^{l-1}$. 
To find an {\em r}-mode for the barotropic case we 
combine this solution
with the numerical solution to (\ref{bspert}). Then we find a 
single mode-solution (that satisfies the boundary conditions)
for a dimensionless eigenvalue $\omega_2$, where
\begin{equation}
 \omega = -{\Omega \over l+1}\left[ (l-1)(l+2) -   
 2\omega_2 { R^3\Omega^2\over M} \right] \ .
\end{equation}
For the identified solution we find that $S_{l+1l}\approx\zeta_{l+1l}$.
In the following we will only consider this specific solution. 
The peculiarity of
the barotropic case will be discussed in more detail elsewhere.
     
It should be noted that the {\em r}-mode frequency
always has different signs in the inertial and the rotating frame. 
That is, although the modes appear retrograde in the rotating system, 
an inertial observer would view them as prograde. 
To leading order, the pattern speed of the modes is
\begin{equation}
\sigma = \Omega { (l-1)(l+2) \over l(l+1) } \ .
\end{equation}
Since $0<\sigma <\Omega$ for all $l\ge 2$, the 
{\em r}-modes are destabilized by the standard CFS mechanism and   
are unstable due to the emission of gravitational waves, 
cf. Eq. (\ref{egw}).
   
We have done a series of calculations for barotropic perturbations of
a polytropic stellar model $p=\kappa \rho^\Gamma$ with $\Gamma=2$. 
The numerical results listed in Table~\ref{tab1} were 
obtained for a model with 
$R=12.47$ km and $M=2.21\ {\rm km} = 1.5 M_\odot$.
The absolute upper limit of rotation is set by the 
the Kepler frequency at which mass shedding at the equator
occurs. This frequency is well approximated by 
$\Omega_K \approx 0.67\sqrt{\pi G\bar{\rho}}$ (where $\bar{\rho}$ is
the average density of the star) for many equations of state. For our
chosen stellar model we have
$\Omega_K \approx 5.6\times10^3\rm\;s^{-1}$, i.e. a minimum period 
$P_K = 2\pi/\Omega_K\approx 1.12$ ms. 

In order to estimate the growth timescale for the {\em r}-mode instability we
need to consider the dissipation due to gravitational waves as
well as the effects of bulk and shear viscosity. To do this we use
 approximations that have been used to study
unstable {\em f}-modes (Ipser and Lindblom 1991).
Specifically, we assume that the    
 time-scale $(t_d)$ associated with
any dissipation mechanism is estimated by
\begin{equation}
{dE \over dt} = -{2E \over t_d} \ .
\end{equation}
Here, $E$  is the
energy of the mode measured 
in the rotating frame. That is, we have (to leading order)
\begin{equation}
E = {1 \over 2} \int \rho |\dot{\xi}|^2 dV  = 
{l(l+1) \over 2} (\omega+l\Omega)^2 \int \rho a^4 |K_{ll}|^2 da \ .
\label{erot}\end{equation}
The dissipation due to gravitational waves can be calculated from the
standard multipole formula (Thorne 1980; Ipser and Lindblom 1991)
\begin{equation}
\left. {dE\over dt}\right|_{\rm gw} = - (\omega +m\Omega) \sum_{l=2}^\infty N_l
\omega^{2l+1} \left( | \delta D_{lm} |^2 + | \delta J_{lm} |^2 \right) \ ,
\label{gwlum}\end{equation}
where 
\begin{equation}
N_l = 4\pi { (l+1)(l+2) \over l(l-1)[(2l+1)!!]^2 } \ .
\end{equation}
The mass multipoles follow from
\begin{equation}
\delta D_{lm} = \int \delta \rho a^l Y_{lm}^* dV \ ,
\label{mass}\end{equation}
and
the current multipoles are given by
\begin{equation}
\delta J_{lm} = 2 \sqrt{ {l\over l+1}} \int a^l (\rho \delta \bar{v} +
\delta \rho \bar{v}) \bar{Y}_{lm}^{B*} dV \ ,
\label{curr}\end{equation} 
where $\bar{Y}_{lm}^{B}$ are the magnetic multipoles defined by Thorne
(1980).
Since we are only interested in the leading order contribution for 
the $l=m$ modes here, the expression for the gravitational wave
luminosity simplifies considerably. Recalling that the Eulerian variation
of the density is second order;  $\delta \rho \sim \Omega^2$, 
and that $\delta \bar{v} = \dot{\bar{\xi}}\sim
\Omega$ we see that the dominant contribution for $l=m$ modes will
come from the first term in (\ref{curr}). Hence, we only need
to evaluate
\begin{equation}
\left. {dE\over dt}\right|_{\rm gw} = - 4 l^2 N_l (\omega +l\Omega)^3 
\omega^{2l+1} \left| \int_0^R \rho a^{l+3} K_{ll} da \right|^2 \ .
\label{egw}\end{equation}
It should be noted that the right-hand side of this equation is positive
for the modes we are discussing.  
This indicates that the emission of gravitational radiation causes
a growth in the mode energy in the rotating frame, despite
the decrease in the inertial frame energy. 
At first sight this may seem puzzling, but it can be understood
from the relation between the two energies:
\begin{equation} 
E_{\rm rot} = E_{\rm inertial} - \Omega J \ .
\end{equation}
From this it is clear that $E_{\rm rot}$ may  increase if 
both $E_{\rm inertial}$  and $J$ decrease (Friedman \& Schutz 1978).
 
The timescale at which the mode-instability grows due to 
gravitational
wave emission is relevant, but it does not in itself establish that we are
dealing with an  astrophysical effect. 
It is well-known that viscosity tends to suppress a
gravitational-wave
instability (Cutler and Lindblom 1987). 
For the instability to be relevant it must 
grow sufficiently fast that it is not completely
damped out by viscosity. For neutron stars, two kinds of viscosity 
--- bulk and shear --- are normally 
considered.
The dissipation due to bulk viscosity follows from 
\begin{equation}
\left. {dE\over dt}\right|_{\rm bv} = - \int \zeta  | \delta \sigma |^2 \ ,
\end{equation}
where 
\begin{equation}
\delta \sigma = -i (\omega+m\Omega) {\Delta p \over \Gamma p} \ .
\label{lagra}\end{equation}
Here $\Delta p$ is the Lagrangian variation in the 
pressure, that can be deduced from (in the notation of Saio (1982)) 
\begin{equation}
\Delta p = \delta p + {dp \over da} \xi^a = \rho g a \sum_{lm} 
\left[ \zeta_{lm} - S_{lm} \right] Y_l^m \ ,
\end{equation}
since $dp/da = -\rho g$ for a Newtonian model. 
The relevant viscosity coefficient is
\begin{equation}
\zeta  = 6\times 10^{25} 
\left( \rho \over 10^{15}{\rm g/cm}^3 \right)^2 \left( {\omega +m\Omega \over
1 {\rm Hz} } \right)^{-2}
\left( { T \over 10^9 {\rm K}} \right)^6 {\rm g/cms}\ .
\end{equation}
After doing the angular integrals we get (for $l=m$)
\begin{equation}
\left. {dE\over dt}\right|_{\rm bv} = 
- \int_0^R \zeta \left( {\rho g a^2 \over \Gamma p} \right)^2
\left[ \zeta_{l+1l} - S_{l+1l} \right]^2 da \ .
\label{bulk}\end{equation}

The dissipation rate due to shear viscosity follows from
\begin{equation}
\left. {dE\over dt}\right|_{\rm sv} = - 2\int \eta \delta\sigma^{ab} \delta 
\sigma^*_{ab} dV \ ,
\end{equation}
where
\begin{equation}
\delta \sigma_{ab} = {{i(\omega+m\Omega)}\over 2} 
\left( \nabla_a \xi_b + \nabla_b\xi_a -2 g_{ab}\nabla_c\xi^c \right) \ .
\end{equation}
This leads to an integral 
\begin{equation}
\left. {dE\over dt}\right|_{\rm sv}
= -(\omega+l\Omega)^2 l(l+1) \left\{  \int_0^R \eta 
 a^2 | a \partial_a K_{ll}|^2 da 
 + (l-1)(l+2) \int_0^R \eta a^2 |K_{ll}|^2 da \right\} \ .
\end{equation}
Above the transition temperature at which the neutron star becomes 
superfluid, the appropriate viscosity
coefficient is
\begin{equation}
\eta  = 2\times 10^{18} 
\left( \rho \over 10^{15} {\rm g/cm}^3 \right)^{9/4} 
\left( { T \over 10^9 {\rm K}}\right)^{-2} {\rm g/cms} \ .
\end{equation}
This estimate should be valid at temperatures
above (say) $10^9$ K, i.e. for adolescent neutron stars.

In Table~\ref{tab1} we list the estimated timescales for the dominant
$l=m=2$ and $l=m=3$ {\em r}-modes. The listed timescales were calculated at
a fiducial temperature $T=10^9$ K.
From the listed data we can deduce the critical rotation period at
which the mode becomes unstable. We do this by solving for the zeros
of
\begin{equation}
{1\over 2E} {dE \over dt} =  -{1\over \tau_{gw}} \left({{\rm 1 ms} \over
P}\right)^{p_{gw}} +  {1\over \tau_{bv}} \left({{\rm 1 ms} \over
P}\right)^{p_{bv}} \left( { 10^9 {\rm K} \over T } \right)^6 + 
{1\over \tau_{sv}}\left( { T \over 10^9 {\rm K}  } \right)^2  \ , 
\label{ebal}\end{equation}
at different temperatures.
The resultant instability window for $l=m=2$ is shown in 
Figure~\ref{fig1}. From the data shown in Figure~\ref{fig1} we can
deduce that the {\em r}-mode instability can potentially limit the star to 
remarkably slow rotation rates. For our specific stellar model
the instability is active for $\Omega/\Omega_K>0.049$, or periods shorter 
than 22.8 ms (at  $T=3.7\times 10^9$K).  
From this it is clear that the {\em r}-mode instability will dominate the
evolution
of a rapidly spinning neutron star.

As we were at the final stage of our calculation
Lindblom, Owen and Morsink (1998) arrived at similar estimates using a 
lowest order approach. For $l=m=2$ they get $\tau_{gw}=18.91(P/{\rm 1 ms})^6$ s
and $\tau_{sv}=2.52\times 10^8$ s, in excellent 
agreement with our results (see Table~\ref{tab1}). The agreement is
not so good in the case of the bulk viscosity. Lindblom et al
estimate that $\tau_{bv}=1.3\times 10^9 (P/{\rm 1ms})^2$ s, i.e. a factor of
70 smaller than our value. 
This surprisingly large difference between the two results
is a consequence of the inaccurate estimate that Lindblom et al use for the 
density perturbation associated with the mode.

To calculate the bulk viscosity, one needs the 
Lagrangian density perturbation.  
But, to leading order in rotation, the density perturbation is zero.  
Lindblom et al work consistently only to leading order, but are able to 
calculate the Eulerian pressure perturbation $\delta p$ to second order 
in rotation directly in terms of their leading-order quantities. They use 
this to compute the Eulerian density perturbation to second order in 
rotation, from the barotropic equation of state.  But the bulk viscosity 
depends on the Lagrangian density perturbation, which can only be 
calculated by fully solving the equations for the perturbation correctly 
to second order in rotation.  We have done that in this paper, and 
we find it makes a larger than expected change in the density perturbation.  
The Lindblom et al approximation amounts to neglecting 
the term $S_{l+1l}$ in the integrand of (\ref{bulk}).  It appears that 
the omission of this term leads to an overestimate 
of the bulk viscosity
dissipation rate, and a low value for $\tau_{bv}$. If we use the same 
approximation (neglect $S_{l+1l}$ in (\ref{bulk})) 
in our calculations we get 
$\tau_{bv}=1.6\times 10^9 (P/{\rm 1ms})^2$ s. This is in good
agreement with the result of Lindblom et al, and proves that our
assertion is correct: In order to arrive at a reliable estimate
for the bulk viscosity one must include both terms in the Lagrangian
variation, as we do in the present paper. That the Lagrangian
variation in the density is much smaller than the
Eulerian one suggests an interesting physical interpretation.
In the rotating star the fluid has moved to remain on the same
(now nonspherical)
surface of constant $\rho$ as in the nonrotating case. This 
is a property worthy of further investigation, but it is not
crucial for our attempt to estimate the limit that the {\em r}-mode
instability sets on the rotation of a young neutron star.
As can be seen in Figure~\ref{fig1}, the limiting rotation speed 
of the star is not a sensitive function of the size of the bulk viscosity, 
essentially because the bulk  viscosity is such a steep function of the 
temperature.  Therefore the large difference between our bulk viscosity 
estimate and that of Lindblom et al (1998) leads to a small 
effective difference in 
the final spin period $P_f$.


\section{Astrophysical implications}

As we have noted in the first section of this paper, there are important 
astrophysical implications of this instability.  It limits the initial 
spin periods of young pulsars, it rules out the accretion-induced collapse 
model for the formation of the fastest millisecond pulsars, and it 
suggests that the formation of neutron stars could be accompanied 
by a strong gravitational wave signal lasting the  cooling time.

We come back to these conclusions below, but we feel it is important
first to assess the influence of various uncertainties 
and approximations on our results.  We do not expect that the qualitative 
picture will change when any of the uncertainties are reduced, but 
the quantitative results for the limiting speed and the spindown timescale 
might be affected.

One source of uncertainty is our simple stellar model. We have used 
a Newtonian polytrope and worked to lowest self-consistent order in 
the spin.  It is important that the problem be considered in the 
framework of general relativity.  As the results of Stergioulas \& Friedman 
(1998) for the $f$-mode instability show,
the critical rotation period (at which the mode becomes unstable)
changes by something like 15\% once general relativity in incorporated in the calculation.
It is possible that the $r$-modes are affected in a similar way. 
Furthermore, a relativistic calculation is required
if we want to investigate issues related to the equation of state.
Within a Newtonian calculation
a comparison of results for different equations of state
is not meaningful since all realistic neutron star equations of 
state are for relativistic models and there is no one-to-one correspondence
with the Newtonian problem. For example, for a given central density
our chosen polytrope $p=\kappa \rho^\Gamma$
will not produce stars  with the same $M$ and $R$ in 
Newtonian theory and general relativity.  Fully relativistic mode 
calculations for rotating stars are very difficult and it may be some years 
before reliable results are obtained from efforts that are only now 
beginning.  But initial investigations within the slow-rotation 
approximation for relativistic stars (Hartle \& Thorne 1968) 
should be possible more quickly.

Another uncertainty affecting our conclusions is the initial 
rotation of the star after collapse.  Clearly, if the initial 
spin of the star is small (as suggested by Spruit and Phinney (1998)), 
then the instability will not operate.
In particular there would be no gravitational radiation.  But 
this seems to us to be unlikely to hold in all cases: the 
angular momentum required to make the star spin at $\Omega_K$ is 
small in the initial degenerate core, and the 16~ms pulsar in N157B is
clear evidence that some pulsars are born spinning rapidly. 

Another important initial condition that we have assumed is uniform 
rotation, but non-axisymmetric collapse is 
unlikely to leave the star in perfect rigid rotation.  If the 
$r$-mode phase is preceded by a dynamical phase involving  short-lived 
bar-mode instabilities (eg, Houser et al 1994 and Rampp et al 1998), 
then the angular velocity field in the 
subsequent axisymmetric star in which the $r$-mode will grow may be highly 
differentially rotating.  This may have some effect on the dynamics 
and growth of the instability (the $r$-modes have not been investigated 
in differentially rotating models), and it may also have an effect 
on the final period.  For example, it could happen that the inner 
and outer regions of the star cool at different rates (see below), 
or that viscosity affects them differently (especially when superfluidity 
sets in).  Then gravitational radiation may extract angular 
momentum from the outer layers, say, spinning them down to near 
$P_f$, but the inner core would remain rapidly rotating.  Once 
the instability ends, the star would eventually become uniformly 
rotating, spinning up the outer layers.  It may be that some such 
process was at work in N157B.

\begin{figure}[htb]
\epsfysize=12cm
\centerline{\epsffile {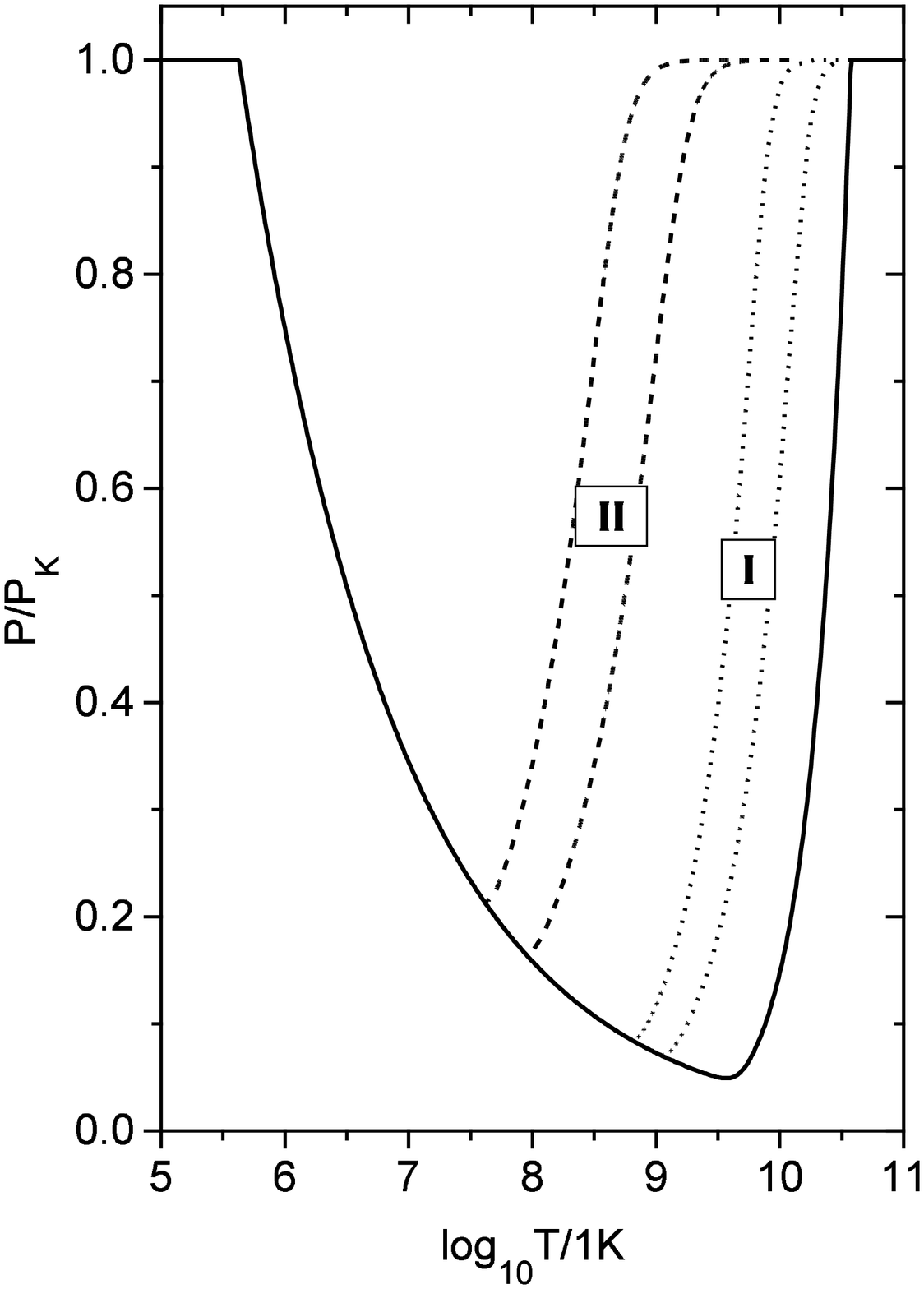}}
\caption{The {\em r}-mode instability window (for $l=m=2$). 
The mode is unstable
above the solid line. Also shown (as dashed lines)
are period evolutions for two different 
cooling scenarios: I) standard cooling due to the modified URCA process, and
II) rapid cooling due to the direct URCA process. In both cases we
assume that the star is isothermal and
 show
two different evolutions corresponding to an {\em r}-mode saturated at 1\% and 100\%
of the 
angular momentum of the star, respectively.}
\label{fig1}\end{figure}

The physics of neutron stars also provides important uncertainties.  Here 
one can hope that observations, coupled with better calculations, can 
provide evidence about this physics.  One possible uncertainty is 
the cooling rate.  
As is clear from Figure~\ref{fig1}, the effect of the $r$-mode instability 
depends largely on the temperature of the star. To assess the astrophysical
role of the instability we must assume a cooling scenario for a 
nascent neutron star. In the standard model, neutron stars are born at 
a temperature well above $10^{10}$ K and cool through the modified URCA
process to $10^9$ K in something like a year. Results obtained
for this model by Nomoto and Tsuruta (1987) are reasonably well
approximated by
\begin{equation}
{T_c \over 10^9 K} = 19.156  \left( {1s \over t} \right)^{0.159} \ , 
\label{slowcool}\end{equation}
where $T_c$ is the central temperature of the star.
Recently, it has been suggested that the star can cool much faster than this.
The idea is that the proton fraction in the star could be sufficently
large to make the direct URCA process possible. Then the star would cool 
according to (Lattimer et al 1994)
\begin{equation}
{T_c \over 10^9 K} = 2.115  \left( {1s \over t} \right)^{0.25} \ . 
\end{equation}
That is, the central parts of the star would cool to $10^9$ K in 20 seconds. 

When combined with the $r$-mode instability, these two 
cooling scenarios will inevitably lead to rather different results. 
To illustrate this, we have made a simple estimate
of how the instability affects the rotation of the star. 
We do this by assuming that
the mode is excited to a maximal level 
corresponding to a certain fraction of the 
total angular momentum of the star (as should be the case 
after an initial time of 
exponential growth, cf. the results of Lai \& Shapiro (1995) 
for the evolution of unstable ellipsoids) 
and that the star loses rotational energy at a 
rate determined by (\ref{ebal}). 
Since we do not know at what level the mode saturates
we consider a wide range of possibilities and allow the mode to 
carry from one percent to all of the stars angular momentum.
Results for this kind of evolutions for our two cooling scenarios are shown
in Figure~\ref{fig1}. In the case of standard cooling (case I in the figure)
the star reaches a final period (the $r$-mode becomes stable) in the range
$13.2 \le P/{\rm 1 ms} \le 15.9$  while in the
rapid cooling case (case II in the figure) 
we get $5.2 \le P/{\rm 1 ms} \le 6.8$.

In these examples we assumed a uniform temperature distribution in the
star.  This is, however, doubtful in the case of rapid cooling.     
When the core of the star cools rapidly, the outer 10\% (say)
of the star will act as a heat reservoir and the star will not be
isothermal until after something like 10 years (cf. Figure~3 of
Lattimer et al (1994)). In effect the temperature of the surface
layers, where the {\em r}-modes are mainly located, is stalled above
$10^9$ K for the first year(s). When this effect is taken into account
the cooling rate relevant for the $r$-modes is closer to that for the 
standard cooling scenario. Hence, we could expect the $r$-mode evolution 
to be close to case I in  Figure~\ref{fig1} even for fast cooling. 
This means that  the instability
would be active for more than two years.

The details of the evolution of the instability are very sensitive 
to calculations of neutron-star viscosity.  The largest 
uncertainty seems to be in the onset of superfluidity.
It is typically assumed that neutron stars become superfluid
at a temperature of the order of $10^9$ K (see Figure~1 of Nomoto \& Tsuruta 
(1987)).
As was shown by Lindblom \& Mendell (1995), the  so-called mutual friction
kills the $f$-mode instability completely in a superfluid star.
This effect is also likely to counteract the $r$-mode instability.
If superfluidity sets in first at the center of the star, it 
could lead to a complex spin evolution, especially if the star is 
differentially rotating.

Within these uncertainties, the simple calculations presented 
here and in Lindblom et al (1998) are consistent with the limited
data that we have on pulsar initial spin rates.  The agreement with 
the presumed initial rotation period of the Crab pulsar of roughly 19 ms 
is extremely good.  If other uncertainties play no role, then this 
would argue that superfluidity did not set in until below $10^9$~K in 
this star.  
The recently discovered 16~ms pulsar associated with the supernova
remnant N157B (Marshall et al (1998)) provides a more stringent test. 
If we assume that the estimated
age of roughly 5000 years is correct and that the pulsar has a braking
index (not yet determined by observations) in the range between those
of the Vela and Crab pulsars ($1.4<n<2.5$) we infer an initial 
period of $6-9$ ms, two to three times as fast as our limit.  This 
suggests either that more accurate physics and model calculations will 
show that $P_f$ is twice as high as our present number (in which 
case the Crab simply formed at a slow initial speed), or that 
initial conditions of formation (eg differential rotation, as 
discussed above) play a big role in determining 
the final spin speed.  It could also be an indication that 
superfluidity played an important role. If
the {\em r}-mode instability were suppressed at $T\approx 2-3\times10^9$ K
by superfluidity our model would agree perfectly with the
observations, cf. Figure~\ref{fig1}.
 Further observations of N157B may provide 
some clues.  In particular it is important to be sure that the 
16~ms period is the true rotation period and not half the true 
period (i.e. that we are not seeing two nearly identical pulses 
per rotation period).

It seems certain that the $r$-mode instability makes it impossible 
to form the fastest millisecond pulsars (at least $P < 10$~ms, 
if not slower) through the accretion-induced collapse of a white dwarf
(see for example Narayan and Popham (1989), Bailyn and Grindlay
(1990) and Nomoto and Kondo (1991)).  Such a star would have to 
form hot, since the collapse is nearly adiabatic.  Even if it 
were spinning rapidly at first, it would have to spin down 
to the kind of speed seen in young pulsars. Subsequent spin up, 
i.e. by continued  accretion, is 
necessary for periods below (say) 10 ms.  Hence, we suggest that all fast 
millisecond pulsars are ``recycled''.

The implications for $r$-mode gravitational radiation are exciting. 
The amount of energy emitted depends essentially on the initial 
spin of the star, but the frequency evolution depends on details 
of the evolution of the star that are beyond the scope of 
this paper.  Owen et al (1998) present a first approximation to 
the evolution of the star as it spins down.  They show that 
the radiation could be detectable by advanced gravitational wave 
detectors.  Observations by such detectors should eventually 
provide new insight into the physics uncertainties we have 
discussed here.

This radiation would be in addition to that which might be 
produced in the initial collapse dynamics.  If the collapsed 
object forms initially rotating faster than the threshold for 
a bar-mode instability (which is near the Kepler period), then 
recent studies by, for example, Houser et al (1994) and 
Rampp et al (1998) indicate that a few percent of the mass 
will be ejected, and that 
the collapsed object will lose angular momentum due to the 
bar-mode instability. The spin will still be sufficiently 
fast that the $r$-mode instability takes over, and a further 
phase of gravitational radiation emission occurs.  

The important astrophysical implications make it 
urgent that progress be made to reduce the uncertainties in 
these calculations.  Fully relativistic calculations, allowing 
for realistic initial conditions and realistic cooling 
scenarios, should lead to results that can confidently 
be compared with radio, X-ray and gravitational-wave observations of 
neutron stars.

\section*{Acknowledgements}

We thank Curt Cutler, John Friedman, Yasufumi Kojima, 
Lee Lindblom and Nick Stergioulas for helpful
discussions. NA and KDK thank the AEI for generous hospitality.

\pagebreak

\begin{table}[h]
\caption{Estimated timescales for the various dissipation mechanisms 
that are relevant to the {\em r}-mode instability. 
The estimates were obtained
by fitting $t= \tau (P/{\rm 1 ms})^p$ s , where $t$ is the e-folding time 
associated with each mechanism and $P$ is the rotation period of the star,
to numerically obtained data.
The gravitational-wave results correspond to index $gw$, while
the bulk viscosity is $bv$ and the shear 
viscosity is $sv$ (notably independent of $P$).   
The timescale data were obtained for a star at $10^9$ K. }
\begin{tabular}[t]{ccccccc}

$l$ & $m$ & $\tau_{gw}$(s) & $p_{gw}$ & $\tau_{bv}$ (s) & $p_{bv}$ & $\tau_{sv}$ (s)  \\  \hline 
&&&&&&\\
2 & 2 & 20.83 & 5.93 & $9.3\times 10^{10}$ & 1.77  & $2.25\times 10^8$ \\
3 & 3 & 316.1 & 7.98 & $1.89\times 10^{10}$ & 1.83 &  $3.53\times
10^7$ \\
\hline
 \end{tabular}
\label{tab1}\end{table}

\pagebreak

\begin{references}
\reference{ligo}
Abramovici A. et al 1992  {\em Science} {\bf 256} 325


\reference{r1} Andersson  N. 1998 {\em A new class of unstable modes of 
rotating relativistic stars} to appear in {\em Ap. J}  [gr-qc/9706075]  

\reference{bg} Bailyn C.D. and Grindlay J.E. 1990 {\em Ap. J.}
{\bf 353} 159

\reference{virgo} Bradaschia C. et al 1990 {\em Nucl. Instr. Meth. Phys. A} 
518

\reference{r2} Chandrasekhar S. 1970  {\em Phys. Rev. Lett.} {\bf 24}
611

\reference{ch} Chandrasekhar S. and Lebovitz N.R. 1962 {\em Ap. J.}
{\bf 132} 1082 

\reference{r3} Cutler C. and Lindblom L 1987 {\em Ap. J.} {\bf 314} 234

\reference{fip} Friedman J.L., Parker L. and Ipser J.R.   1989
{\em Phys. Rev. Lett. } {\bf 62} 3015

\reference{r4} Friedman J.L. and Morsink S. 1998 
{\em Axial instability of rotating relativistic stars}
to appear in {\em Ap. J.} [gr-qc/9706073]

\reference{r5}  Friedman J.L. and Schutz  B.F. 1978 {\em Ap. J.} {\bf 222}
281

\reference{r6} Glendenning N.K. 1996 {\em Compact Stars} (Springer Verlag,
New York)

\reference{HARTLE1968} Hartle J.B. and Thorne K.S. 1968 {\em Ap. J.} {\bf 153} 
807

\reference{house} Houser J.L., Centrella J.M. and Smith S.C. 
1994 {\em Phys. Rev. Lett.} {\bf 72} 1314  


\reference{r7} Ipser J.R.  and Lindblom L. 1991  {\em Ap. J.} {\bf 373} 213 

\reference{ls} Lai D. and Shapiro S.L. 1995 {\em Ap. J.} {\bf 442} 259
\reference{r8}  Lattimer J.M.,  van Riper K.A., Prakash M. and  Prakash M. 1994
{\em Ap. J.} {\bf 425} 802 

\reference{r9} Lindblom L. 1995 {\em Ap. J.} {\bf 438} 265 
 
\reference{r10} Lindblom L. and Mendell G.  1995 {\em Ap. J.} {\bf 444} 804

\reference{r11} Lindblom L., Owen B.J. and Morsink S.M. 1998 {\em Gravitational
radiation instability in hot young neutron stars} to appear
in {\em Phys. Rev. Lett.}  [gr-qc/9803053]

\reference{mm} Marshall F.E., Gotthelf E.V., Zhang W., Middleditch
J. and Wang Q.D. 1998 {\em Discovery of an Ultra-fast pulsar in the
supernova remnant N157B} [astro-ph/9803214] 

\reference{mtw} Misner C.W., Thorne K.S. and Wheeler J.A. 1973 {\em
Gravitation} (W.H. Freeman: San Francisco)   

\reference{np} Narayan R. and Popham R. 1989 {\em Ap. J. Lett.} {\bf
346} 25

\reference{nk} Nomoto K. and Kondo Y. 1991 {\em Ap. J. Lett.} {\bf
367} 19

\reference{r12} Nomoto K. and Tsuruta S. 1987 {\em Ap. J. } {\bf 312} 711 

\reference{ow} Owen B, Lindblom L., Cutler C., Schutz B.F., Vecchio A. and
Andersson N. 1998
{\em Gravitational waves from hot young rapidly rotating neutron
stars} to appear in {\em Phys. Rev. D} [gr-qc/9804044]

  
\reference{r13} Papaloizou J. and  Pringle J.E. 1978 {\em MNRAS} {\bf
182} 423


\reference{r14} Provost J., Berthomieu G. and Rocca A. 1981 
{\em Astron. Astrop.} 
{\bf 94} 126

\reference{retal} Rampp M., M\"uller E. and Ruffert M. 
1998 {\em Astron. Astrop.} {\bf 332} 969

\reference{r15} Saio H.  1982  {\em Ap. J.} {\bf 256} 717

\reference{sch} Schutz B.F. 1983 {\em Lect. Appl. Math.} {\bf 20} 99 

\reference{phin} Spruit H. and Phinney E.S. 1998 {\em Why pulsars
rotate and move: kicks at birth} [astro-ph/9803201]

\reference{r16} Stergioulas N. and  Friedman J.L. 1998  {\em Ap. J.} 
{\bf 492} 301

\reference{r17} Thorne K.S. 1980 {\em Rev. Mod. Phys.} {\bf 52} 299  

\reference{r18} Thorne K.S. 1987 in {300 years of gravitation}  ed:
S.W. Hawking and W. Israel p. 330 (Cambridge Univ. Press)

\end{references}
\end{document}